\documentclass[final,authoryear,5p]{elsarticle}

\biboptions{authoryear}
\usepackage{natbib}
\usepackage{epsfig}
\usepackage{float}
\usepackage[export]{adjustbox}

\usepackage{amssymb}
\usepackage{amsmath}

\usepackage[colorlinks = true, linkcolor = blue, urlcolor  = blue, citecolor = blue, anchorcolor = blue]{hyperref}
\usepackage{lineno}
\journal{Advances in Space Research}
\begin{document}

\begin{frontmatter}

\title{Quasi-biennial oscillations in the cross-correlation of properties of macrospicules}

\author[label1,label2]{T. S. Kiss\corref{cor}}
\cortext[cor]{Corresponding author}
\ead{tskiss1@sheffield.ac.uk}

\author[label1,label3]{N. Gyenge}
\author[label1]{R. Erd\'elyi}
\address[label1]{Solar Physics and Space Plasmas Research Centre (SP2RC), School of Mathematics and Statistics, University of Sheffield,\\ Hicks Building, Hounsfield Road, Sheffield, S3 7RH, United Kingdom}
\address[label2]{Department of Physics, University of Debrecen, Egyetem t\'er 1., Debrecen, H-4032, Hungary}
\address[label3]{Debrecen Heliophysical Observatory (DHO), Konkoly Astronomical Institute, Research Centre for Astronomy and Earth Sciences\\Hungarian Academy of Sciences, Debrecen, P.O.Box 30, H-4010, Hungary}

\begin{abstract}

Jets, whatever small (e.g. spicules) or large (e.g. macrospicules) their size, may play a key role in momentum and energy transport from photosphere to chromosphere and at least to the low corona. Here, we investigate the properties of abundant, large-scale dynamic jets observable in the solar atmosphere: the macrospicules (MS). These jets are observationally more distinct phenomena than their little, and perhaps more ubiquitous, cousins, the spicules. Investigation of long-term variation of the properties of macrospicules may help to a better understanding of their underlying physics of generation and role in coronal heating. Taking advantage of the high temporal and spatial resolution of the Solar Dynamics Observatory, a new dataset, with several hundreds of macrospicules, was constructed encompassing a period of observations over six years. Here, we analyse the measured properties and relations between these properties of macrospicules as function of time during the observed time interval. We found that cross-correlations of several of these macrospicule properties display a strong oscillatory pattern. Next, wavelet analysis is used to provide more detailed information about the temporal behaviour of the various properties of MS. For coronal hole macrospicules, a significant peak is found at around 2-year period. This peak also exists partially or is shifted to longer period, in the case of quiet Sun macrospicules. These observed findings may be rooted in the underlying mechanism generating the solar magnetic field, i.e. the global solar dynamo.

\end{abstract}


\end{frontmatter}

\parindent=0.5 cm
\section{Introduction}
Since their first detection \citep{bohlin1975}, evolution and generation of macrospicules (MS) have been investigated in several ways. These jets appear and disappear at around 20-40 minutes timescales and their length and width could be multiple times that of the spicules. The notion of spicules and macrospicules may refer to a possible relation between the two phenomena, however, their behaviour, generation and evolution seem all to be rather different.

Majority of MS studies are based on imaging observations. The early works focus on determining the physical dimensions of the observed jets in intensity maps. The first 25 MS were identified by the spectroheliograph onboard \textit{Skylab} \citep{bohlin1975}. These results were followed up quickly by the analysis of another two macrospicules and their energy was estimated, by the application of a cylindrical assumption of their geometry to be: $3 \times 10^{26}$ ergs \citep{withbroe1976}.
 
Statistical analysis of 32 MS was carried out by \cite{labonte1979}. They found MS jets using various wavelengths such as H$\alpha$ and D$_{\text{3}}$ at Big Bear Observatory. \cite{dere1989} also investigated 10 MS of \textit{Skylab-2} observations at EUV wavelenghts. 

In these early works, macrospicules could be divided into two subclasses based on the wavelength of the observation: the EUV macrospicules and H$\alpha$ macrospicules. \cite{shibata1982} filled in the gap between the two classes: the author claimed that the main difference between the two classes of jets is the intensity ratio between the macrospicule and the surrounding corona environment.

Improvement of observation technology led to a more complex analysis of these jets. Other physical properties, e.g. temperature and density, became measurable, therefore plenty of new features of MS were found: i) it has been shown that MS jets are multi-thermal phenomena, which have the potential to be the source of the fast solar wind \citep{pike1997}; ii) emission of the two opposite sides along the main axis of the jets are blue and redshifted, therefore MS may be rotating objects ("solar tornadoes") \citep{pike1998}; iii) the density of MS is around $10^{10}$ cm$^{-3}$ and their temperature is about 2--3 $\times 10^{5}$ K \citep{parenti2002}, and iv) the energy needed for MS formation could be between 3.66 $\times 10^{13}$ J -- 1.46 $\times 10^{17}$ J \citep{bennett2015}.

Numerical simulations of MS were developed side by side with the observations. The first numerical simulation of macrospicules was carried out by \cite{murawski2011}. The authors claimed that chromospheric velocity pulses may have the ability to generate macrospicules.

Furthermore, \cite{loboda2017} compared observations taken by the TESIS solar telescope and from simulation results of an axially symmetric one-dimensional hydrodynamic method. They found that macrospicules loose $\approx$ 12\% of their original mass during their lifetime. 
 
Recently, some works claimed to find a possible connection between the properties of MS and the behaviour of the global solar dynamo. For example \cite{gyenge2015} investigated the spatial distibrution of 101 MS between June 2010 and December 2012. 
Their longitudinal and latitudinal distribution showed a qualitatively similar pattern, what could be also observed in the case of spatial distribution of the sunspot groups.

Long-term investigation of MS was carried out by \cite{kiss2017}. In their work, basic physical properties of MS (such as length, width, area, lifetime, upflowing velocity) were studied over a 5-year period of observation. A strong oscillatory pattern was discovered with around a two-year period. Particular study of this wave signature was really difficult due to the short time interval of MS selection. The length of operation time of \textit{Solar Dynamics Observatory (SDO)} \citep{pesnell2012} and its \textit{Atmospheric Imaging Assembly (AIA)} \citep{lemen2012} limited the opportunities to observe jets continuously for a longer period of time, therefore MS are searched for between 01.06.2010 and 31.12.2015. Five years of observation was just not sufficiently long enough for signal processing methods to provide high-enough confidence information about the oscillation.

Therefore the aim of this paper is to extend the database, investigate these oscillation signatures and provide more confidently  detailed information about them.
\begin{figure*}[t]
	\includegraphics[scale=0.22]{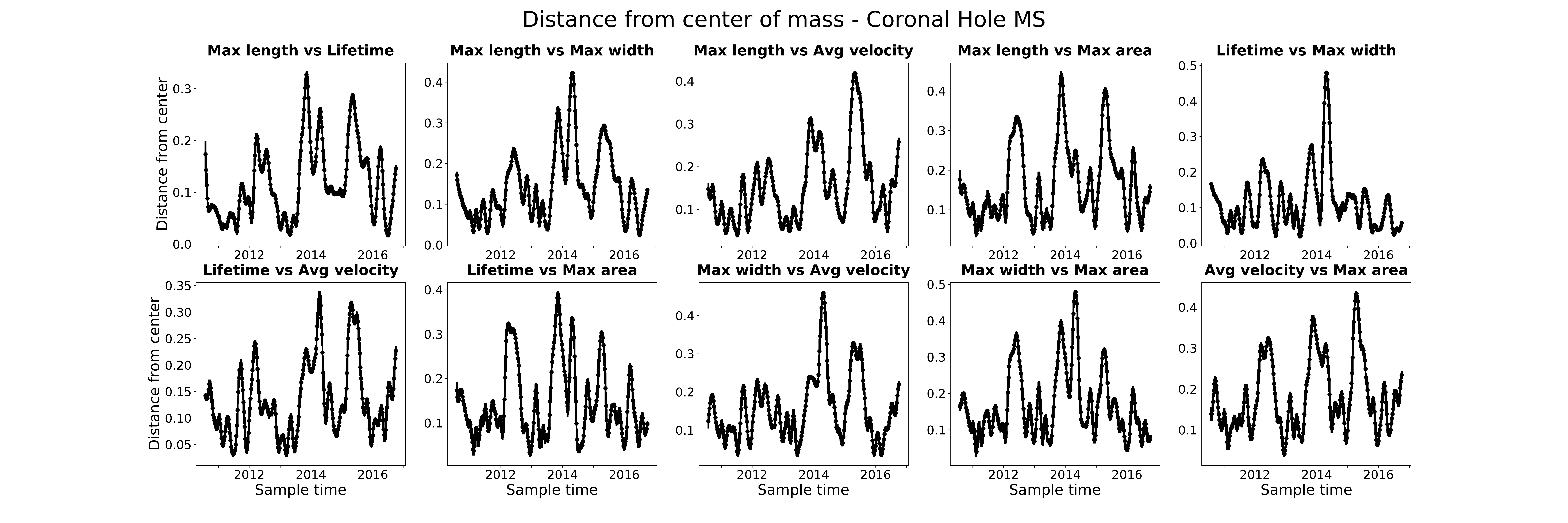}\\
	\includegraphics[scale=0.22]{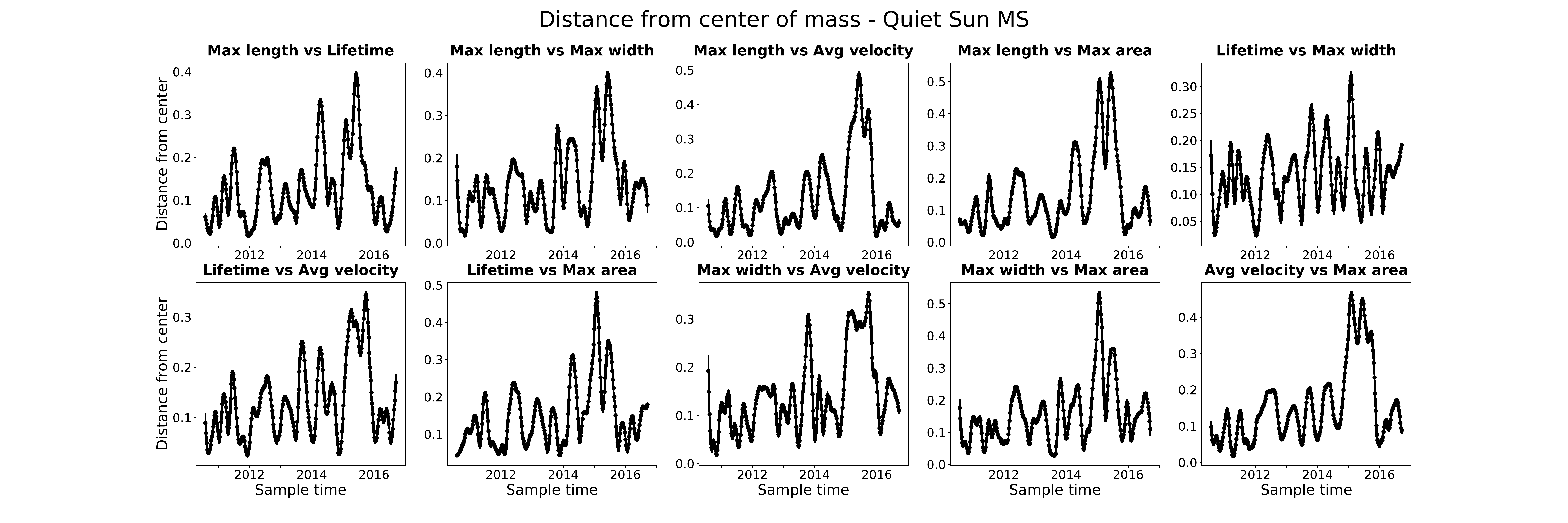}
	\centering
	\caption{Diagrams of the geometric distance from the center of the mass for a range of cross-correlated physical parameters. Variation for CH-MS are in the top two rows, while the bottom two rows contain temporal information about QS-MS.}
	\label{fig01}
\end{figure*}

\begin{figure*}[t]
	\includegraphics[scale=0.22]{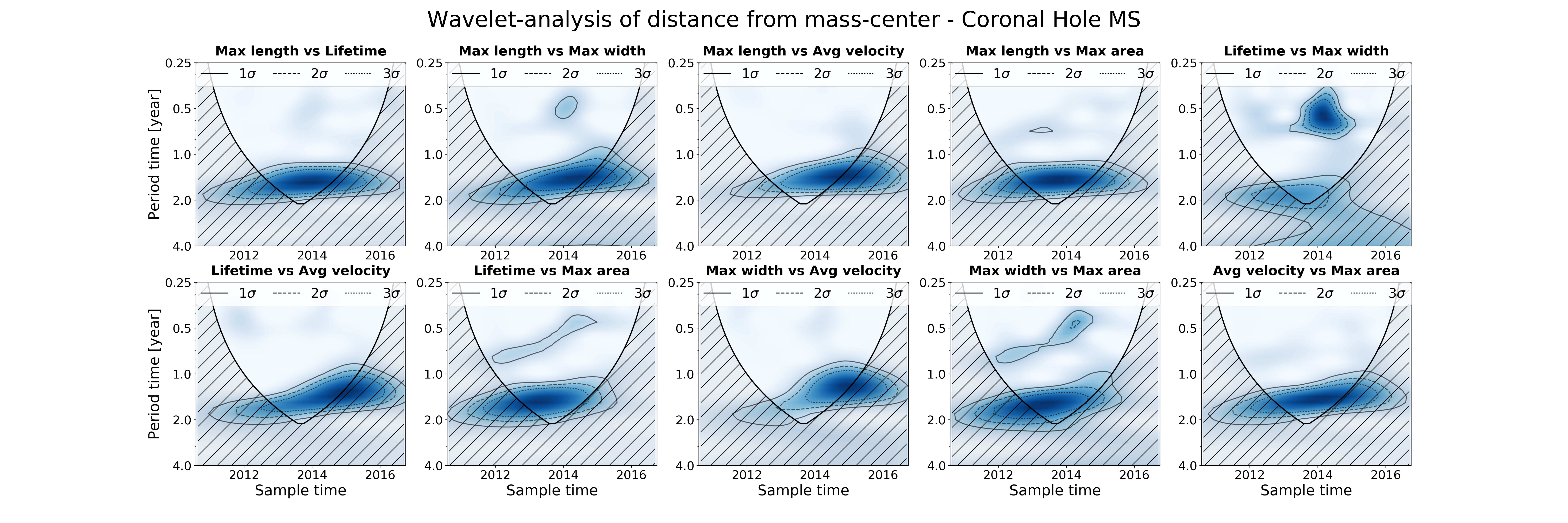}\\
	\includegraphics[scale=0.22]{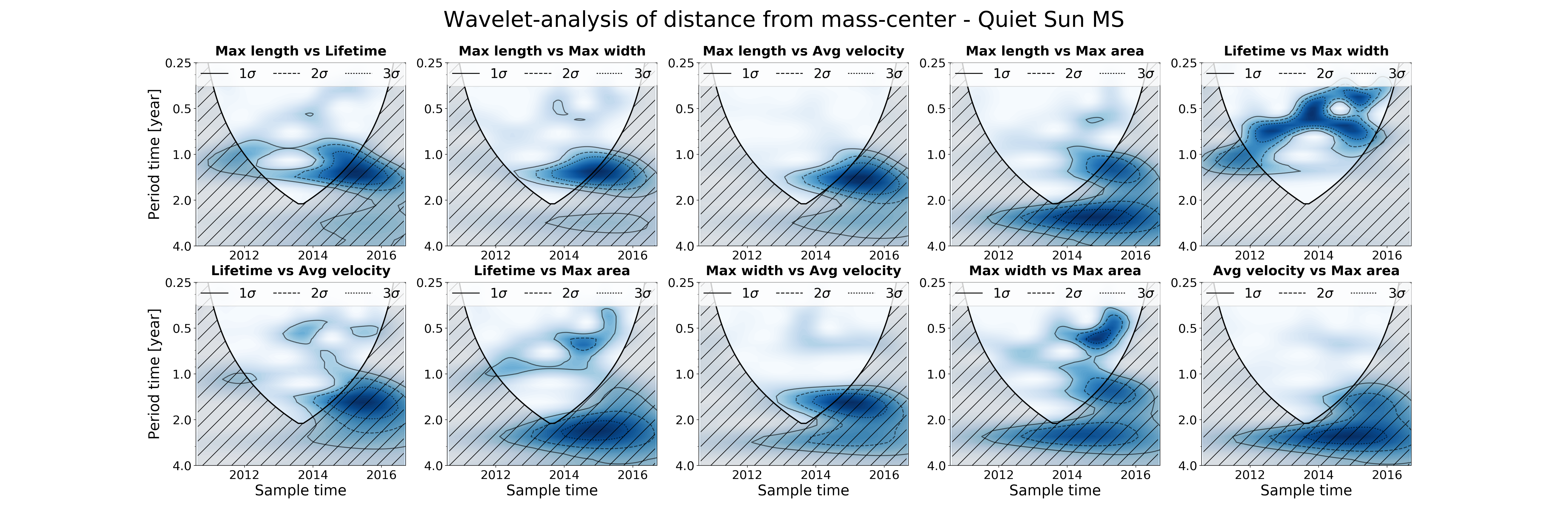}
	\centering
	\caption{Wavelet analysis of the distance from the center of mass for cross-correlated physical properties of MS. The upper two rows are of CH-MS, the bottom two rows of QS-MS. Solid, dashed and dotted contour lines represent the $1\sigma, 2\sigma$ and $3\sigma$ deviations from the average. The COI is outlined by a black line and grey filling.}
	\label{fig02}
\end{figure*}

\section{Database and Methodology}
Long-term investigation of MS properties requires a temporally homogeneous database. To achieve this goal, the data provider instrument should meet several requirements such as operation time for multiple years and adequate temporal resolution for resolving MS evolution. SDO was launched in 2010 and its AIA instrument carries out a full-disc image of the Sun at nine different wavelengths. For the database, the 30.4 nm, optically thin, EUV line was used, where MS are clearly visible. Temporal resolution of \textit{SDO/AIA} in 30.4 nm is 12 s, therefore the second criteria listed above is fulfilled. During the research, we used Python and its solar data analysis package, the Sunpy \citep{mumfort2015}.

Detailed information about the criteria of how to define MS and the so-called tetragon assumption are all described in \cite{kiss2017}, we would not repeat them here. However, let us briefly recall the key steps for completeness. We searched MS in a 2-hour long observating window on a given day. As the log-normal distribution fitting of several MS property shows in \cite{kiss2017}, the characteristic lifetime of MS is significantly shorter ($\approx$ 15--16 min) than the temporal length of the observing window. Therefore the entire evolution of these jets can be observed comfortably during these 2 hours. For this reason, \textit{SDO/AIA} 30.4 nm observations between 12:00--14:00 on each 1$^{\text{st}}$, 7$^{\text{th}}$, 15$^{\text{th}}$ and 24$^{\text{th}}$ day of each month since June 2010 (launch of \textit{SDO}) until the end of 2016 are the basis of the data analysis. An advantage of this choice of data selection is that the sampling is homogeneous. This is a required criteria for the application of wavelet analysis.
 
The application of set of definitions yields the discovery of 342 MS. The data were collected on 312 observing dates, therefore an average of $\approx 1.1$ MS are visible during a two-hour interval at the solar limb in 30.4 nm. Furthermore, the tetragon model assumption provides a rather accurate description of the physical properties (such as area, length, width) of a jet for each \textit{SDO/AIA} observation during MS lifetime. This set of macrospicules is a representative sample.

The maximum values of these properties (maximum length, maximum width, maximum area) in addition with average velocity and lifetime are cross-correlated with each other. Considering the temporal evolution of these distributions, strong oscillatory patterns become visible. This wavy signature was found first in the distance data between the actual cross-correlation point and the center of the mass, if a geometric approximation is used for these plots. In \cite{kiss2017}, there is a complete description of such analysis. 

Signal processing methods are powerful tools to gain detailed information about subtle features present during oscillations. However, these applications often require temporally homogeneous database. The previously described selecting method provides nearly homogeneous on multiple years timescale, since the length of temporal grid between two neighbouring dates is $6 \pm 1$ days. This variance can be neglected due to the comparison to the length of the database. However, occurrence rate of MS on each date is not similar at all: the average number of MS on each observation date is 1.1, but often varies between null and three (highest number of MS on one day is five). This variance could cause an unwanted influence when applying signal processing methods, such as wavelet- or Fourier-analysis. To solve this problem, a smoothing method was constructed, which averages out missing MS and complements the the dates without MS observation by using linear interpolation.

To obtain information about the periodic behaviour in the distance data, we used wavelet analysis. The wavelet analysis is an appropriate tool to find different frequencies with non-stationary power in the evolution of a given signal \citep{daubechies1990}. The reason behind using wavelet analysis rather than Fourier analysis is to catch the temporal variation of the oscillation peaks. For more details, see, \cite{torrence1998} providing a practical description of application of wavelet analysis through meteorological examples.

In this paper, an extended and temporally homogeneous database is used. One more year of observation yields another 43 MS (23 Coronal Hole macrospicules [CH-MS] and 20 Quiet Sun macrospicules [QS-MS]) and grows the observational window up to 6.5 years. This time interval now has the required length to detect oscillations more confidently with periods around 1.5-2 years outside the cone of influence (COI) defined in the wavelet.

\section{Results}
The distance of each cross-correlation point from the center of mass is plotted in Figure~\ref{fig01}, separately for Coronal Hole and Quiet Sun MS. The two plots (CH-MS and QS-MS) seem to show characteristic differences, which may root in the different physical properties of the jets. 

For a more complete investigation of these features, wavelet-analysis was applied, and the results are shown in Figure~\ref{fig02}. For CH-MS, a strong peak is visible between 1-2 years. This peak is only weaker in the case of the cross-correlation "Lifetime vs Maximum width" panel, however, an overtone appears there at around 0.5 year. This overtone exists with significance in further three cross-correlation plots (e.g., "Maximum length vs Maximum width", "Lifetime vs Maximum area", "Maximum width vs Maximum area") and with less power in the other cases. 

Analysis of the QS-MS geometric distance shows a more complex evolutionary property. The oscillation with period between 1-2 years exist, but mostly only after 2014. However, their domain is inside the COI, thus more observation is needed to confirm its significance. The source of this period is a wave peak at around the first part of 2015, which shows up for almost each QS-MS cross-correlation. Further, a 3-year long periodicity also can be identified in the analysis, with significance for "Maximum width vs Maximum area", "Lifetime and Maximum width", "Lifetime vs Maximum area" and, furthermore, less significantly for nearly all cases. The most interesting case is the cross-correlation "Lifetime vs Maximum width", where the 0.5-year long oscillation is the dominant one with a less powerful 1-2-year period peak. In several cases, a strong peak appears during a 2-4 years long domain, but these regions are in the COI.

\section{Conclusions and discussion}
This study focuses on proving further evidence for and, most importantly, confirming of the existence of the conjectured oscillations with nearly a two-year period time reported in the distance plots of \cite{kiss2017}. Temporal length of that dataset was not sufficiently long enough to identify with high confidence the multiple-year long oscillatory patterns. Therefore, even a one-year extension of this dataset leads to an opportunity to investigate the evolution of physical properties of macrospicules, in a statistical sense, to result in adequately higher confidence levels. A wavelet analysis is applied separately to the data of Coronal Hole and Quiet Sun MS. Detailed investigation of these wavelets are the main results of this paper. To avoid spurious effects in the wavelet analysis, the recent more extended datasets is homogenised temporally. This step is an important one in order enable us to have a more accurate estimate. Almost every cross-correlated data shows a significant peak period of oscillation. In the case of CH-MS, a strong peak takes place between one and two years and a less significant peak at around 0.5-year. We recall that several other oscillation periods around two-year period (called Quasi-Biennial Oscillations -- QBOs) were already reported in previous studies. Even more interestingly, \cite{beaudoin2016} constructed a new generation solar dynamo: this modelling generates two dynamo layers. One dynamo is at the bottom of the convective zone and produces the 11-year long solar cycle. The second dynamo layer (may) work(s) near the solar surface and may be able to generate QBOs around the solar poles. As shown in \cite{kiss2017}, CH-MS are mostly formed around the solar poles. The cross-correlations shown in this paper are applied to the various measured physical properties of MS properties, so the question still stands: what can be the source of the reported periodicities? Upcoming further research will aim to solve this problem.

For QS-MS, the peak period of around 1-2 year is mostly within the COI and appears after 2013-2014. This feature may refer to a change in the source of MS generation. Driver of several chromospheric jets, i.e. macrospicules, may be a small-scale, local magnetic field, which emerges into a previously existed magnetic field \citep{sterling2000}. It is worth recalling that maxima of the 24th solar cycle occurred around 2013-2014, when the oscillation in the cross-correlation of properties of QS-MS become significant. The overtone of around 0.5-year appears in these cases, but with more power, than in the case of CH-MS.

If the signature of an overtone would be confirmed with other observations, that may further support the concept of a double-dynamo operating in the Sun.

Advantage of wavelet analysis is, among others, the easy visualization of the evolution of the oscillation peaks. In several cases, this variation is strongly visible. Motivation of future research could be to understand, why the period of QBOs varies as the solar cycle progresses.

This investigation serves as an example towards demonstrating the potential of this continuously growing dataset we have constructed about MS. Even with a one-year addition of MS observation to the data analysed by \cite{kiss2017}, here we were able to confirm the existence of QBOs in some key physical properties of MS conjectured in \cite{kiss2017}. Further continuation of the data gathering during the solar cycle would be essential in order to establish whether there are longer periods  present, and if so, are they linked to the cyclic variations itself.

\section{Acknowledgement}
The authors thanks the support received from the Erasmus Programme of EU and the ESPRC (UK) for supporting this research.  RE is grateful to STFC (UK), The Royal Society and the Chinese Academy of Sciences Presidents International Fellowship Initiative, Grant No. 2016VMA045 for support received.

\end{document}